\documentclass[a4paper]{article}

\usepackage{graphicx}
\usepackage[]{color}
\usepackage{amsmath,latexsym}
\usepackage{setspace}

\makeatletter
\def\maxwidth{ %
  \ifdim\Gin@nat@width>\linewidth
    \linewidth
  \else
    \Gin@nat@width
  \fi
}
 {\par\unskip\endMakeFramed%
 \at@end@of@kframe}
\makeatother

\definecolor{fgcolor}{rgb}{0.345, 0.345, 0.345}

\usepackage{framed}

\definecolor{shadecolor}{rgb}{.97, .97, .97}
\definecolor{messagecolor}{rgb}{0, 0, 0}
\definecolor{warningcolor}{rgb}{1, 0, 1}
\definecolor{errorcolor}{rgb}{1, 0, 0}

\makeatletter
\newcommand{\@affiliations}{}
\newcommand{\affiliations}[1]{\renewcommand{\@affiliations}{#1}}
\newcommand{\@running}{}
\newcommand{\running}[1]{\renewcommand{\@running}{#1}}
\newcommand{\@nwords}{}
\newcommand{\nwords}[1]{\renewcommand{\@nwords}{#1}}

\renewcommand\maketitle{\begin{titlepage}%
    \let\footnotesize\small
    \let\footnoterule\relax
    \let \footnote \thanks
    \singlespacing%
    \textbf{Running title:} \@running \par%
    \textbf{Number of words:} $\sim$\@nwords \par%
    \vskip 1em%
    \textbf{Date of submission:} \@date \par%
    \vskip 4em%
    \begin{center}%
      {\Large \sffamily \bfseries \@title \par}%
      \vskip 3em%
      {\large \sffamily
        \lineskip .75em%
        \begin{tabular}[t]{c}%
          \@author
        \end{tabular}\par}%
    \end{center}\par
    \vskip 4em%
    \begin{enumerate}%
      \itshape%
      \@affiliations%
    \end{enumerate}%
    \@thanks
    \vfil\null
  \end{titlepage}%
  \setcounter{footnote}{0}%
  \global\let\thanks\relax
  \global\let\maketitle\relax
  \global\let\@thanks\@empty
  \global\let\@author\@empty
  \global\let\@date\@empty
  \global\let\@title\@empty
  \global\let\title\relax
  \global\let\author\relax
  \global\let\date\relax
  \global\let\and\relax
}

\renewenvironment{abstract}{%
  \small
  \begin{center}%
    {\sf\bfseries \abstractname\vspace{-.5em}\vspace{\z@}}%
  \end{center}%
  \quotation}
{\endquotation}

\renewcommand\section{\@startsection {section}{1}{\z@}%
  {-3.5ex \@plus -1ex \@minus -.2ex}%
  {2.3ex \@plus.2ex}%
  {\Large\sf\bfseries}}
\renewcommand\subsection{\@startsection{subsection}{2}{\z@}%
  {-3.25ex\@plus -1ex \@minus -.2ex}%
  {1.5ex \@plus .2ex}%
  {\large\sf\bfseries}}
\renewcommand\subsubsection{\@startsection{subsubsection}{3}{\z@}%
  {-3.25ex\@plus -1ex \@minus -.2ex}%
  {1.5ex \@plus .2ex}%
  {\normalfont\normalsize\sf\bfseries}}
\makeatother

\usepackage{alltt}
\usepackage{graphicx}
\usepackage[english,francais]{babel}
\usepackage{fancyhdr}
\usepackage{lastpage}
\usepackage{natbib}
\usepackage{url}
\usepackage{pdfcolmk}
\usepackage[colorlinks=true,colorlinks=true,%
           citecolor=blue,%
           filecolor=blue,%
           linkcolor=blue,%
           urlcolor=blue]{hyperref}

\newcommand{\pa}[1]{\left(#1\right)}
\newcommand{\ac}[1]{\left\{#1\right\}}

\newcommand{\tN}{\widetilde{N}}
\newcommand{\hN}{\widehat{N}}
\newcommand{\tE}{\widetilde{E}}
\newcommand{\hE}{\widehat{E}}
\newcommand{\tP}{\widetilde{P}}

\newcommand{\V}{\mathcal{V}}
\newcommand{\tk}{\widetilde{k}}

\nwords{8600}

\begin{document}


\title{Capitalizing on Opportunistic Data for Monitoring Species Relative Abundances}

\running{Capitalizing on opportunistic data}

\author{Christophe Giraud, Cl\'{e}ment Calenge, Camille Coron \& Romain Julliard}

\affiliations{
\item C. Giraud, 
  Laboratoire de Math\'ematiques d'Orsay, UMR 8628, Universit\'e
  Paris-Sud, France and Centre de Math\'ematiques Appliqu\'ees, UMR 7641, Ecole Polytechnique, France.
\item C. Calenge (\textnormal{\url{clement.calenge@oncfs.gouv.fr}}), 
  Office national de la chasse et de la faune sauvage, Direction des
  \'etudes et de la recherche, Saint Benoist, BP 20. 78612 Le Perray en
  Yvelines, France.
\item C. Coron, 
  Laboratoire de Math\'ematiques d'Orsay, UMR 8628, Universit\'e
  Paris-Sud, France.  
\item R. Julliard, CESCO, UMR 7204, MNHN-CNRS-UPMC, CP51, 55 rue
  Buffon, 75005 Paris, France.
}




\label{firstpage}

\maketitle
\begin{abstract}
  With the internet, a massive amount of information on species
  abundance can be collected under citizen science programs. However,
  these data are often difficult to use directly in statistical
  inference, as their collection is generally opportunistic, and the
  distribution of the sampling effort is often not known.  In this
  paper, we develop a general statistical framework to combine such
  ``opportunistic data'' with data collected using schemes
  characterized by a known sampling effort. Under some structural
  assumptions regarding the sampling effort and detectability, our
  approach allows to estimate the relative abundance of several
  species in different sites. It can be implemented through a simple
  generalized linear model. We illustrate the framework with
  typical bird datasets from the Aquitaine region, south-western
  France.  We show that, under some assumptions, our approach provides
  estimates that are more precise than the ones obtained from the
  dataset with a known sampling effort alone. When the opportunistic data are abundant, the gain in precision
  may be considerable, especially for the rare species. We also
  show that estimates can be obtained even for species recorded only
  in the opportunistic scheme.  Opportunistic data combined with a
  relatively small amount of data collected with a known effort may
  thus provide access to accurate and precise estimates of
  quantitative changes in relative abundance over space and/or
  time.
\end{abstract}



\newpage

\section{Introduction}

How species abundance varies in space and time is a major issue both
for basic (biogeography, macroecology) and applied (production of
biodiversity state indicators) ecology. Professionals working on
biodiversity thus spend considerable resources collecting data that
are suitable for estimating this variation \citep{Yoccoz2001}.  Most
of the scientific literature recommends the implementation of both a
statistically valid sampling design and a standardized protocol for
collecting such data \citep[e.g. see][for a
review]{Williams2002a}. Many methods have been developed to estimate
species abundance in a defined location, e.g., using mark-recapture
methods \citep{Seber1982} or distance sampling approaches
\citep{Buckland1993}. However, these approaches require an intense
sampling effort and are not always practical. Many authors have noted
that most frequently, interest will not be in abundance itself, but
either in the rate of population change, i.e., the ratio of abundance
in the same location at two different time points, or in the relative
abundance, i.e., the ratio of abundance at two separate locations
\citep{MacKenzie2002a}.

Relative abundance is frequently monitored with the help of simpler
schemes. For instance, a set of sites is randomly sampled in the area
of interest, and counts of organisms are organized on these sites
using a given protocol. At a given location, the resulting count can
be used as an index of the true abundance. Indeed, assuming constant
detectability over space and time, the average number of animals
counted per sampled site is proportional to the true abundance of the
species in the area. Log-linear models can be used to represent this
average number of animals detected per site as a function of space
and/or time \citep[and, possibly, other factors such as the habitat;
see for example][]{Strien2001}, and thereby, to infer population
trends. Thus, such programs have been implemented in many countries to
monitor the changes in the abundance of several groups of species,
such as birds \citep[e.g., for the French Breeding Bird Survey,
see][]{Julliard2004} or butterflies \citep[e.g., for the European
Butterfly Monitoring Scheme, see][]{Swaay2008}. Estimates of
relative abundance have also been commonly used for mapping the
spatial distribution of several species \citep[e.g.,][]{Gibbons2007}.

In addition to such data characterized by a known sampling effort, a
large amount of data can also be collected by non-standardized means,
with no sampling design and no standardized protocol. In particular,
the distribution of the observers and of their sampling effort is
often unknown \citep{Dickinson2010}. These so-called ``opportunistic
data'' have always existed, and with the recent development of citizen
science programs, we observe a massive increase in the collection of
these data on a growing number of species \citep[e.g.,
][]{Dickinson2010,Hochachka2012,Dickinson2012}. Additionally, as the
use of online databases facilitates the exchange and storage of data,
such opportunistic data may now include millions of new observations
per year that are collected in areas covering hundreds of thousands of
square kilometres \citep[e.g., the global biodiversity information
facility, including more than 500 million records at the time of
writing, see][]{Yesson2007}.

The temporal and spatial distributions of the observations in such
data reflect unknown distributions of both observational efforts and
biodiversity. Thus, a report of a high number of individuals of a
given species at a given location compared to other locations could be
because the focus species is abundant at this location or because
numerous observers were present at this location. Using such
opportunistic data to estimate changes in the space and time of
species abundance is therefore complex, since any modeling approach
should include a submodel of the observation process
\citep{Kery2009,Hochachka2012} or an attempt to manipulate the data to
remove the bias caused by unequal effort \citep[see a discussion
in][]{Phillips2009}.

As noted by \citet{MacKenzie2005}, ``In some situations, it may be
appropriate to share or borrow information about population parameters
for rare species from multiple data sources. The general concept is
that by combining the data, where appropriate, more accurate
estimates of the parameters may be obtained.'' In this paper, we
propose a general framework which enables to combine data with known
observational effort (which we call ``standardized'' data) with
``opportunistic'' data with an unknown sampling effort. We focus on
multi-species and multi-site data that correspond to the data
typically collected in this context.

The purpose of this study is to estimate the relative abundance of the
species at different sites (different locations and/or times). We base
this estimation on two datasets recording the number of animals
detected by observers for each species of a pool of species of
interest and each spatial unit of a study area of interest: (i) one
``standardized'' dataset is collected under a program characterized by
a known sampling effort, possibly varying among spatial units, (ii)
one ``opportunistic'' dataset is characterized by a completely unknown
sampling effort. We take into account the variation across species of
their detectability, yet, as a first step, we assume that the
observational bias towards some species are the same across the
different sites.  We show that, under this assumption, the information
concerning both the distribution of the observational effort and the
biodiversity can be efficiently retrieved from ``opportunistic'' data
by combining them with standardized data. Moreover, we prove that such
a combination returns more accurate estimates than when using the
standardized data alone. Our statistical framework allowing this
win-win combination can open numerous avenues for application. We
used data on French birds, which are typical of existing data, to
illustrate the numerous qualities of this framework. Note however that
the work presented in this paper is a first step, and that further
work will be required to fully account for varying observational bias
towards some habitat types across the different sites.

During the reviewing process of this paper, we became aware of an independent and simultaneous work by \citet{Fithian2014} which develops similar ideas for 
combining multi-species and multi-sites data with thinned Poisson models. 

\section{Statistical modeling}

We want to estimate the relative abundance (relative number of
individuals) of $I$ species in $J$ sites.  The ``sites'' $j$ can
either refer to different spatial sites, to different times, or to
different combinations of sites and times.  We suppose that we have
access to $K$ datasets indexed by $k$ which gather counts for each
species $i$ at each site $j$. We have in mind a case where some
datasets have been collected with some standardized protocol, while
some others are of opportunistic nature.

Let $X_{ijk}$ be the count of individuals of the species $i$ by the
observers in the site $j$ in the dataset $k$.
In this paper, we propose to model the counts $X_{ijk}$ by
\begin{equation}\label{model}
X_{ijk}\sim \textrm{Poisson}(N_{ij}P_{ik}E_{jk}),\ \ \textrm{for } i=1,\ldots,I,\ j=1,\ldots,J\ \textrm{and } k=0,\ldots,K-1,
\end{equation} 
where $N_{ij}$ is the number of individuals (animals, plants, etc) of
a species $i$ at site $j$, and $P_{ik}$, $E_{jk}$ are two parameters
accounting for the bias induced by the observational processes.  The
parameter $P_{ik}$ reflects both the detectability of the species $i$
(some species are more conspicuous than others, some are more easily
trapped, etc.) and the detection/reporting rate of this species in the dataset
$k$ (the attention of the observers may systematically vary among
species).  The parameter $E_{jk}$ reflects the impact of the varying
observational effort (including number and duration of visits, number
of traps, etc.) and the varying observational conditions met during
the counting sessions.  In the next two sections, we explain the
origin of our modeling, the hypotheses under which it is valid (see
also the discussion Section~\ref{sec:discussion}), and we describe precisely the meaning of the two
dimensionless parameters $P_{ik}$ and $E_{jk}$.  We refer to the
Appendix~\ref{sec:thinned} for a discussion on the link with models
based on thinned Poisson processes.  Before moving to these modeling
issues, we point out that estimation can be easily carried out in the
model~(\ref{model}), since it can be recast into a linear generalized
model, see Section~\ref{sec:estimation}.

\subsection{Count modeling}\label{sec:count:modeling}

The count $X_{ijk}$ of individuals of the species $i$ in the site $j$
for the dataset $k$ is assumed to gather the counts from all visits in
the site $j$.  We assume that an individual is only counted \emph{once
  during a single visit}, yet it can be counted \emph{several times}
in any dataset due to the possible \emph{multiple visits} to a site
$j$ for a dataset $k$.  In particular, we may have $X_{ijk}$ larger
than the number $N_{ij}$ of individuals of the species $i$ in the site
$j$. In the following, we neglect identification errors and false
positives.

For an individual $a_{ij}$ of the species $i$ in the site $j$ and a
visit $v_{jk}$ in the site $j$ for the dataset $k$, we define the
random variable $Z_{a_{ij}v_{jk}}$ which equals 1 if the individual
$a_{ij}$ has been \emph{seen and recorded} during the visit $v_{jk}$,
and 0 otherwise. Assuming that there is no multiple count of an
individual during a single visit, the count $X_{ijk}$ is then given by
\begin{equation*}
X_{ijk}=\sum_{v_{jk}\in\V_{jk}} \sum_{a_{ij}=1}^{N_{ij}}Z_{a_{ij}v_{jk}},
\end{equation*}
where $\V_{jk}$ is the set of all the visits $v_{jk}$ in the site $j$
for the dataset $k$.  In the following, we denote by
$p_{a_{ij}v_{jk}}=\mathbf{P}\pa{Z_{a_{ij}v_{jk}}=1}$ the probability
for the individual $a_{ij}$ to be seen and recorded during the visit
$v_{jk}$.  \medskip

If we assume that the random variables $\{Z_{a_{ij}v_{jk}}:a_{ij}=1,\ldots,N_{ij}\ \textrm{and}\ v_{jk}\in\V_{jk}\}$ are independent and that
\[ \sum_{v_{jk}\in\V_{jk}}\sum_{a_{ij}=1}^{N_{ij}}p_{a_{ij}v_{jk}}^2\ \ \textrm{is small compared to}\ \ \sum_{v_{jk}\in\V_{jk}}\sum_{a_{ij}=1}^{N_{ij}}p_{a_{ij}v_{jk}},\]
(which happens when $p_{a_{ij}v_{jk}}$ is small),
then, according to Le Cam Inequality~\citep{lecam},  the count $X_{ijk}$ follows approximatively the Poisson distribution 
\begin{equation}\label{eq:model1}
X_{ijk}\sim\textrm{Poisson}\bigg(\sum_{v_{jk}\in\V_{jk}}\sum_{a_{ij}=1}^{N_{ij}}p_{a_{ij}v_{jk}}\bigg)=\textrm{Poisson}\bigg(N_{ij}\sum_{v_{jk}\in\V_{jk}}\bar p_{iv_{jk}}\bigg),\quad \textrm{with}\ \ \bar p_{iv_{jk}}={1\over N_{ij}}\sum_{a_{ij}=1}^{N_{ij}}p_{a_{ij}v_{jk}}.
\end{equation}
The parameter $\bar p_{iv_{jk}}$ corresponds to the average probability to detect and report during the visit $v_{jk}$ an individual of the species $i$ which has been sampled at random in the site $j$. 
We observe that the mean of the Poisson distribution
\[N_{ij}\sum_{v_{jk}\in\V_{jk}}\bar p_{iv_{jk}}=N_{ij}O_{ijk}\]
is the product of a first term $N_{ij}$, which is the number of individuals of the species $i$ present in the site $j$, by a second term $O_{ijk}$, which is a nuisance term due to the observational process. We underline that the term $O_{ijk}$ can be larger than 1 when the number  $V_{jk}$ of visits in the site $j$ for the dataset $k$ is large, since an individual can be counted several times during the $V_{jk}$ visits.

\subsection{Main modeling assumption}\label{sec:assumptions}
The main hypothesis of our modeling~(\ref{model}) is that the observational parameter $O_{ijk}$ can be decomposed as
\begin{equation}\label{assume:main}
O_{ijk}=P_{ik}E_{jk}.
\end{equation}
Let us give three examples where such a decomposition holds.

\noindent{\bf Example 1. (single habitat type)}  Assume that the ratios $\bar p_{iv_{jk}}/\bar p_{i'v_{jk}}$ depend only on the species $i$ and $i'$ and on the dataset $k$, so that  $\bar p_{iv_{jk}}/\bar p_{i'v_{jk}}=\bar p_{iv'_{j'k}}/\bar p_{i'v'_{j'k}}$ for all $i,i',j,j',v_{jk}$, and $v'_{j'k}$. This means that the 
detection/reporting probability $\bar p_{iv_{jk}}$ of an individual of the species $i$ during the visit $v_{jk}$
 can be decomposed as 
\begin{equation}\label{example1}
\bar p_{iv_{jk}}=P_{ik}q_{v_{jk}},
\end{equation}
with
 $P_{ik}$ the mean detection/reporting probability of the species $i$ during a visit for the dataset $k$ and
 $q_{v_{jk}}$ depending only on the visit $v_{jk}$ (not on the species $i$). 
The parameter $q_{v_{jk}}$ represents the influence of the observational conditions during the visit $v_{jk}$ on the detection/reporting probability. The parameter $q_{v_{jk}}$ is then a very complex function of the observational duration, the visibility conditions (weather conditions during the visit, vegetation met, etc.)  and many other variables that affect the detection/reporting probability (number of traps, length of line transects, etc.). 
When the decomposition~(\ref{example1}) holds, we have the decomposition~(\ref{assume:main}) with $E_{jk}=\sum_{v_{jk}\in\V_{jk}} q_{v_{jk}}$. 

The decomposition (\ref{example1})  enforces that the detection/reporting probability $\bar p_{iv_{jk}}$ does not depend on interactions between the species $i$ and the visit $v_{jk}$. This property is quite restrictive and it is not likely to be met when several habitat types are present within a site $j$. Actually, if two visits $v_{jk}$ and $v'_{j'k}$ take place in two different habitat types $h_{jk}$ and $h'_{j'k}$ then the ratios $\bar p_{iv_{jk}}/\bar p_{i'v_{jk}}$ and $\bar p_{iv'_{j'k}}/\bar p_{i'v'_{j'k}}$ are not likely to be equal for all $i$ and $i'$ since some species may be specialized to the habitat type $h_{jk}$ and some others to the habitat type $h'_{j'k}$. We can weaken the assumption (\ref{example1}) by allowing interactions $\epsilon_{iv_{jk}}$ between the species $i$ and the visit $v_{jk}$ as long as they cancel on average on each site $j$
\begin{equation}\label{example1bis}
\bar p_{iv_{jk}}=P_{ik}q_{v_{jk}}+\epsilon_{iv_{jk}},\ \ \textrm{with}\ \ \sum_{v_{jk}\in\mathcal{V}_{jk}}\epsilon_{iv_{jk}}\simeq 0.
\end{equation}
When (\ref{example1bis}) holds, we again have the decomposition~(\ref{assume:main})  with $E_{jk}=\sum_{v_{jk}\in\V_{jk}} q_{v_{jk}}$. Such interactions $\epsilon_{iv_{jk}}$ can take into account heterogeneous observer attention bias toward the species $i$, but it does not allow for some systematic bias induced by heterogeneous habitat types. 
Actually, assume that the site $j$ has two habitats $h$ and $h'$ and the site $j'$ has only the habitat $h'$. Then if the species $i$ (respectively $i'$) is specialized to habitat type $h$ (respectively $h'$) we will have either $\sum_{v_{j'k}}\epsilon_{i v_{j'k}}<0$  or $\sum_{v_{jk}}\epsilon_{i' v_{jk}}<0$. So (\ref{example1bis}) cannot hold. The next two examples focus on  the impact of heterogeneous habitat types.

\noindent{\bf Example 2. (known habitat types)} 
Assume that for each count, we know in which habitat type it has occurred. 
Let us introduce the parameter $\tk=(h,k)$ where $h$ represents the habitat type $h$ and $k$ the dataset. For each dataset $k$, we can then pool together the counts occurring in the same site $j$ and habitat type $h$. Let us denote by $X_{ij(h,k)}$ the counts of the species $i$ in the site $j$, the habitat type $h$ for the dataset $k$. 
We assume in the following that each visit occurs in a single habitat type: If not, we can artificially split a single visit in $H$ different habitat types into $H$ different visits, each occurring in a single habitat type. 

Our main modeling assumption in this example is that the
 ratios $\bar p_{iv_{j(h,k)}}/\bar p_{i'v_{j(h,k)}}$ depend only on the species $i$ and $i'$, the dataset $k$ and the habitat type $h$. This means that for each $i,i',j,j'$ and $\tk=(h,k)$ we have
$\bar p_{iv_{j\tk}}/\bar p_{i'v_{j\tk}}=\bar p_{iv'_{j'\tk}}/\bar p_{i'v'_{j'\tk}}$ for  all visits $v_{j\tk},v'_{j'\tk}$ in the same dataset and the same habitat type.
In this case, the
 probability $\bar p_{iv_{j(h,k)}}$ can be decomposed as 
\begin{equation}\label{example2}
\bar p_{iv_{j(h,k)}}=P_{i(h,k)}\,q_{v_{j(h,k)}},
\end{equation}
with $P_{i(h,k)}$  the mean detection/reporting probability of a typical individual of the species $i$ during a visit in the habitat type $h$ for the dataset $k$ and $q_{v_{j(h,k)}}$ not depending on $i$.
We then have for $\tk=(h,k)$
\[O_{ij\tk}=P_{i\tk}E_{j\tk},\quad\textrm{with}\ \ E_{j\tk}=\sum_{v_{j\tk}\in\V_{j\tk}}q_{v_{j\tk}}\ \ \textrm{and}\ \ P_{i\tk}=P_{i(h,k)}\ \ \textrm{defined by (\ref{example2}).}\]
As above, we can allow some non-systematic heterogeneity by merely assuming that 
$\bar p_{iv_{j(h,k)}}=P_{i(h,k)}\,q_{v_{j(h,k)}}+\epsilon_{iv_{j(h,k)}}$ with $\sum_{v_{j(h,k)}\in\mathcal{V}_{j(h,k)}}\epsilon_{iv_{j(h,k)}}\simeq 0$.

\noindent{\bf Example 3. (homogeneous habitat type proportions)} We assume again that each visit $v_{jk}$ occurs in a single  habitat type $h(v_{jk})$ (by artificially splitting non-homogeneous visits). Yet, we assume that this habitat type is not reported in the dataset. 
As in the second example, we also assume that the
ratios $\bar p_{iv_{jk}}/\bar p_{i'v_{jk}}$ depend only on the species $i$ and $i'$, the dataset $k$ and the habitat type $h(v_{jk})$. 
Hence, the
 probability $\bar p_{iv_{jk}}$ can be decomposed as 
\begin{equation}\label{example3}
\bar p_{iv_{jk}}=P_{ih(v_{jk})k}\,q_{v_{jk}},
\end{equation}
with $P_{ihk}$  the mean detection/reporting probability of a typical individual of the species $i$ during a visit in the habitat type $h$ for the dataset $k$ and $q_{v_{jk}}$ not depending on $i$. Writing $\mathcal{V}_{jk}(h)$ for the set of the visits $v_{jk}$ in the habitat type $h$ we have
\[O_{ijk}=\sum_{h=1}^H \sum_{v_{jk}\in \mathcal{V}_{jk}(h)} \bar p_{iv_{jk}}= \sum_{h=1}^H P_{ihk} E_{jhk},\quad \textrm{with}\ \   E_{jhk}=\sum_{v_{jk}\in \mathcal{V}_{jk}(h)}q_{v_{jk}}.\]
The parameters $E_{jhk}$ are  likely to depend on $h$ since there can be some observational bias towards some habitat types. If we assume that the  observational bias is the same for each site $j$, which means that $E_{jhk}/E_{j'hk}$ does not depend on $h$, we have the decomposition
\begin{equation}\label{example2bis}
E_{jhk}=E_{jk}Q_{hk},
\end{equation}
where $Q_{hk}$ reflects the observational bias towards the habitat type $h$ in the dataset $k$.
When the decompositions (\ref{example3}) and (\ref{example2bis}) hold, we have
\[O_{ijk}=\sum_{h=1}^HP_{ihk}Q_{hk}E_{jk}=P_{ik}E_{jk},\quad\textrm{with}\ \ P_{ik}=\sum_{h=1}^HP_{ihk}Q_{hk},\]
so $O_{ijk}$ fulfills   the decomposition~(\ref{assume:main}). 
Again, as in the two first examples, we can weaken  (\ref{example3}) by merely assuming that
\[\bar p_{iv_{jk}}=P_{ih(v_{jk})k}\,q_{v_{jk}}+\epsilon_{iv_{jk}},\ \ \textrm{with}\ \ \sum_{v_{jk}\in\mathcal{V}_{jk}}\epsilon_{iv_{jk}}\simeq 0.\]
Let us explore when the decompositions~(\ref{example3}) and~(\ref{example2bis}) are likely to hold. We first observe that 
the decomposition~(\ref{example3}) will be met as long as we include in the definition of the "habitat type" $h(v_{jk})$ all the exogenous variables which induces an interaction between the species $i$ and the visit $v_{jk}$. The decomposition~(\ref{example2bis}) is much more stringent. It requires that, for each dataset $k$, the observational bias towards some habitat types is the same across the different site $j$. It may not hold when the proportions on habitat types differ among the different sites. For example, if an habitat $h$ is missing in a site $j$, then $E_{jhk}=0$, so~(\ref{example2bis}) cannot hold if $E_{j'hk}\neq 0$ for another site $j'$. An example where this property is more likely to be met is when the "sites" $j$ correspond to the same spatial unit observed at different years $j$. In such a case, we can expect that the observational bias towards some habitat types remains stable years after years. 
When the observational bias towards some habitat types is not constant across the site, the decomposition~(\ref{assume:main}) is not met in general. This case requires a substantial additional modeling that will be developed elsewhere.

\noindent{\bf Interpretation.}
Let us interpret more precisely the parameters $P_{ik}$ and $E_{jk}$ in the decomposition~(\ref{assume:main}). Writing $V_{jk}$ for the number of visits in the site $j$ for the dataset $k$, we first observe that 
\[{1\over J}\sum_{j=1}^J{1\over V_{jk}}\sum_{v_{jk}\in\V_{jk}}\bar p_{iv_{jk}}={1\over J}\sum_{j=1}^J{O_{ijk}\over V_{jk}}={1\over J}\sum_{j=1}^JP_{ik}\,{E_{jk}\over V_{jk}}=P_{ik}\bar E_{k},\quad\textrm{with}\ \bar E_{k}=J^{-1}\sum_{j=1}^JE_{jk}/V_{jk}.\]
We can always replace  $(P_{ik},E_{jk})$ in the decomposition~(\ref{assume:main}) by $(P'_{ik},E'_{jk})=(P_{ik}\bar E_{k},E_{jk}/\bar E_{k})$.
Applying this renormalization step and dropping the prime (for notational simplicity),  we obtain
\begin{equation}\label{eq:Pik}
P_{ik}={1\over J}\sum_{j=1}^J{1\over V_{jk}}\sum_{v_{jk}\in\V_{jk}}\bar p_{iv_{jk}},
\end{equation}
which means that $P_{ik}$ is the mean detection/reporting probability of a typical individual of the species $i$ during a typical visit for the dataset $k$.

As explained in the three above examples, 
the parameter $E_{jk}$ in~(\ref{assume:main}) is a complex function of the conditions met during the visits in the site $j$ for the dataset $k$, including the observational effort. 
This parameter $E_{jk}$ can be (much) larger than 1 when the number $V_{jk}$ of visits in the site $j$ for the dataset $k$ is very large.
We point out that we can have $E_{jk}$ very large even if $O_{ijk}$ is smaller than 1, when the probability $P_{ik}$ of detection/reporting of a typical individual of the species $i$ is very small.
In the remaining of the paper, we call \emph{observational intensity}  at the site $j$ in the dataset $k$ the parameter $E_{jk}$.

\subsection{Identifiability issues}

In the following, we deal with two datasets. A first dataset labeled
by $k=0$, in which we suppose that the observational intensities
$E_{j0}$ are known up to a constant. Henceforth, we will call this
dataset the \emph{standardized dataset}. We also consider a second
dataset labeled by $k=1$, characterized by unknown observational
intensities $E_{j1}$. We will refer to this dataset as the
\textit{opportunistic dataset}.

\subsubsection{A single opportunistic dataset is not
  enough}\label{sec:single}

We consider first the case where we have a single dataset, i.e.\
$K=1$. For notational simplicity, we drop the index $k$ in this
paragraph. Our observations $X_{ij}$ then follows a Poisson
distribution with intensity $\lambda_{ij}$, where
$\lambda_{ij}=N_{ij}P_{i}E_{j}$.  We cannot recover the $IJ+I+J$
parameters $N_{ij}$, $P_{i}$, and $E_{j}$ from the $IJ$ intensities
$\lambda_{ij}$.  Yet, if we are only interested by the relative
abundances $N_{ij}/N_{ij'}$ with respect to a reference site, say
$j'=1$, can we recover the $I(J-1)$ ratios
$\ac{N_{ij}/N_{i1}:j=2,\ldots,J,\ i=1,\ldots,I}$ from the $IJ$
parameters $\lambda_{ij}$?

Let us write $\lambda_{ij}=\tN_{ij}\tP_{i}\tE_{j}$ with
$\tN_{ij}=N_{ij}P_{i}E_{1}$ , $\tP_{i}=1$ and $\tE_{j}=E_{j}/E_{1}$.
The parameters $\tN_{ij}$ differ from the $N_{ij}$ by a multiplicative
constant $P_{i}E_{1}$ depending only on the species $i$. Therefore, we
have $N_{ij}/N_{ij'}=\tN_{ij}/\tN_{ij'}$, which means that the
parameters $\tN_{ij}$ give access to the relative abundances
$N_{ij}/N_{i1}$ of the species $i$.  When the dataset has been
collected with a known sampling design, the observational intensity in
a given site $E_{j}$ is known up to an unknown constant, so that the
ratios $E_{j}/E_{j'}$ are known and we can recover the $\tN_{ij}$ (and
hence the relative abundances) from $\lambda_{ij}$ since the $\tP_{i}$
and $\tE_{j}$ are known.  The situation is different with
opportunistic datasets characterized by unknown ratios
$E_{j}/E_{j'}$. In this case, the $\tE_{j}$ are also unknown, so we
cannot recover the $\tN_{ij}$ from the parameters
$\lambda_{ij}$. Hence, we do not have access to the relative abundance
$N_{ij}/N_{i1}$. As explained in the next paragraph, we need to
combine different datasets.

\subsubsection{Combining an opportunistic dataset with a standardized one}
Let us now investigate the identifiability issues when we combine a standardized dataset (labeled by $k=0$) with an opportunistic one (labeled by $k=1$). 
In this case, we have $2IJ$ parameters $\lambda_{ijk}=N_{ij}P_{ik}E_{jk}$ for $IJ+2(I+J)$
parameters $N_{ij}$, $P_{ik}$ and $E_{jk}$. For $IJ>2(I+J)$, which typically holds for large $J$ and
$I\geq 3$, we have more parameters $\lambda_{ijk}$ than parameters $N_{ij}$, $P_{ik}$ and $E_{jk}$. Nevertheless, as
explained in the Appendix~\ref{appendix:proof}, the model is not identifiable
without $J+I+1$ additional identifiability conditions. As in
Section~\ref{sec:single}, we introduce some renormalisation
$\widetilde N_{ij}$, $\widetilde E_{jk}$ of $\widetilde P_{ik}$ of
$N_{ij}$, $E_{jk}$ and $P_{ik}$, which enables us to easily express these
identifiability conditions while preserving the identity  $\widetilde
N_{ij} \widetilde E_{jk} \widetilde P_{ik}=\lambda_{ijk}=N_{ij}E_{jk}P_{ik}$. 

In the following, we assume that the ratios $\ac{E_{jk}/E_{j'k}:j\neq j'}$ are known for the dataset $k=0$ (standardized dataset), but not for the dataset $k=1$ (opportunistic one). As above, we
define $\widetilde E_{j0}=E_{j0}/E_{10}$ (which is known) and $\widetilde
P_{i1}=1$ for all $i$. We could have set $\widetilde P_{i0}=1$ instead
of $\widetilde P_{i1}=1$, but the latter choice is more suited for
handling species $i$ monitored in the dataset $k=1$ but not  in the
dataset $k=0$, as we will show later. We must still set one
more constraint. We choose $\widetilde P_{10}=1$ for
convenience. These $I+J+1$ constraints combined with the identity
$\widetilde N_{ij} \widetilde E_{jk} \widetilde
P_{ik}=\lambda_{ijk}=N_{ij}E_{jk}P_{ik}$ lead to the change of variables:
\begin{eqnarray}\label{renormalization}
\widetilde N_{ij}&=&N_{ij}P_{i1}E_{10}{P_{10}\over P_{11}},\nonumber\\
\widetilde E_{jk}&=&{E_{jk}\over E_{10}}\times {P_{1k}\over P_{10}}\\
\widetilde P_{ik}&=&{P_{ik} \over P_{i1}}\times{P_{11} \over P_{1k}}. \nonumber
\end{eqnarray}
In terms of these new variables, we have the simple statistical model
$X_{ijk}\sim\textrm{Poisson}(\widetilde N_{ij}\widetilde
E_{jk}\widetilde P_{ik})$ with
$\widetilde E_{j0}=E_{j0}/E_{10}$ for all $j$, $\widetilde P_{i1}=1$ for all
$i$ and $\widetilde P_{10}=1$. These $J+I+1$ quantities are
 known, and the resulting statistical model is identifiable.

Let us interpret these new quantities. The parameter $\widetilde
N_{ij}$ is proportional to the abundance $N_{ij}$ by an unknown factor
$P_{i1}E_{10}P_{10}/P_{11}$ depending only on the species $i$. As in 
Section~\ref{sec:single}, these parameters give access to the relative abundance 
$N_{ij}/N_{i1}=\tN_{ij}/\tN_{i1}$ of
each species $i$ in each site $j$. The parameters $\widetilde
E_{j1}$ are equal, up to a constant, to the observational intensity $E_{j1}$; therefore, they provide the relative observational intensities $E_{j1}/E_{11}$ for each site $j$ in the dataset
1. Finally, $\widetilde P_{i0}$ is proportional to the ratio
$P_{i0}/P_{i1}$ by an unknown factor $P_{11}/P_{10}$, so we can
compare the ratios $P_{i0}/P_{i1}$ across the different species. The
ratio $P_{i0}/P_{i1}$ reflects the systematic difference of attention
toward some species among the observers of the two schemes. 

In addition, we emphasize that we can consider the case where some
species $i$ are not monitored in the dataset 0 but are recorded in the
dataset 1. This case can be handled by merely adding the constraints
$\widetilde P_{i0}=P_{i0}=0$ for the concerned species $i$. 

\subsection{Estimation via a Generalized Linear Model}\label{sec:estimation}
We can estimate the parameters $\widetilde N_{ij}$, $\widetilde
E_{jk}$ and $\widetilde P_{ik}$ by the maximum likelihood estimators
$(\widehat N_{ij},\widehat E_{jk},\widehat P_{ik})$ with the
constraints $\widehat E_{j0}=\widetilde E_{j0}$ for all $j$, $\widehat
P_{i1}=1$ for all $i$ and $\widehat P_{10}=1$. This estimation can be
carried out with the help of a generalized linear model. Indeed, with
the notations $n_{ij}=\log(\tN_{ij})$, $e_{jk}=\log(\tE_{jk})$ and
$p_{ik}=\log(\tP_{ik})$, Model~(\ref{model}) can be recast as
a classical generalized linear model from the Poisson family with a
log link:
\begin{equation}\label{GLM}
X_{ijk}\sim\textrm{Poisson}(\lambda_{ijk}),\quad\textrm{with}\
\log(\lambda_{ijk})=n_{ij}+e_{jk}+p_{ik}.
\end{equation}
Indeed, we only have to define $e_{j0} = \log \tE_{j0}$ as a known offset in
the model, $p_{i1} = 0$ for all $i$, and fit the resulting model with
any statistical package (see Supplementary materials).

\section{Theoretical gain of combining two
  datasets} 
  
It is important to investigate whether the estimates of the relative abundance obtained by combining the dataset 1 with
unknown observational intensity ratios $E_{j1}/E_{j'1}$ to the dataset 0 with known observational intensity ratio $E_{j0}/E_{j'0}$ improves
upon the
estimates obtained with the single dataset 0. 
In this section, we investigate this issue analytically.
An improvement is
expected simply by looking at the balance between the number of
observations and the number of free parameters. With the dataset 0, we
have $IJ$ observations, and we want to estimate $IJ$ free parameters;
whereas with the two datasets 0 and 1, we have $2IJ$ observations for
$IJ+J+I-1$ free parameters. The balance between the number of
observations and the number of free parameters is better in the second
case. Below, we quantify the theoretical  improvement more precisely by comparing the variance of the maximum-likelihood estimators in the two cases. Then, we show that dataset combination also allows to estimate relative abundance for species $i$  not monitored in the dataset 0.  
  
\subsection{Variance reduction}  
\label{sec:vared}

For mathematical simplicity, we assume in the following that the ratios
$P_{i0}/P_{i1}$ are known for all $i$. In terms of the normalized
variables, this means that the $\tP_{i0}$ are known.  

When we work
with the single dataset 0, we can estimate $\tN_{ij}$ with the maximum
likelihood estimator $\widehat N_{ij}^0=X_{ij0}/(\tE_{j0}\tP_{i0})$.
Let us investigate how the maximum likelihood estimator $\hN_{ij}$
associated with the model $X_{ijk}\sim\textrm{Poisson}(\widetilde
N_{ij}\widetilde E_{jk}\widetilde P_{ik})$ improves upon $\widehat
N_{ij}^0$. We consider the case where the (unknown) observational intensities $E_{j1}$ in
the dataset 1 is much larger than the observational intensities $E_{j0}$ in the dataset
0.  Hence, we consider the asymptotic setting where $E_{j1}$ goes to infinity.
 In the Appendix~\ref{appendix:proof}, we show that the limit variance of $\widehat
N_{ij}$ when $E_{j1}\to\infty$ is given by
\begin{equation}\label{variance}
\textrm{var}(\widehat N_{ij})\stackrel{E_{j1}\to\infty}{\to}\textrm{var}(\widehat N_{ij}^0)\times {P_{i0}N_{ij}\over
  \sum_{l}P_{l0}N_{lj}}\,.
\end{equation}
In particular, the variance of the estimate is reduced by a factor
\[{\textrm{var}(\widehat N_{ij})\over \textrm{var}(\widehat N_{ij}^0)}
\stackrel{E_{j1}\to\infty}{\approx}  {P_{i0}N_{ij}\over
  \sum_{l}P_{l0}N_{lj}}\,,\]
when working with the two datasets instead of the sole dataset 0. This factor can be very small for rare species ($N_{ij}$ small), hardly detectable species ($P_{i0}$ small), or when the number $I$ of monitored species is large.
 
Let us explain the origin of this variance reduction in the simple case where the ratios $P_{i0}/P_{i1}$ are the same for
all the species $i$ (which formally corresponds to $\widetilde
P_{i0}=1$ for all $i$). In this case, we have a closed-form formula for $\hN_{ij}$
(see Formula~(\ref{MLEclosedformula}) in the Appendix~\ref{appendix:proof})
\[\widehat N_{ij}={X_{ij0}+X_{ij1}\over
  \sum_{l}(X_{ij0}+X_{lj1})}\times {\sum_{l}X_{lj0}\over
  \tE_{j0}}\,,\] which reveals the contribution of each dataset
to the estimation of the (normalized) relative abundance. Actually, the estimator
$\widehat N_{ij}$ is the product of two terms, where the first term
mainly depends on the opportunistic dataset~1 when the 
observational intensities $E_{j1}$ are large, whereas the second term only depends on the
dataset 0
\[\widehat N_{ij}\ \stackrel{E_{j1}\to\infty}{\approx}\ {X_{ij1}\over
  \sum_{l}X_{lj1}}\times {\sum_{l}X_{lj0}\over \tE_{j0}}\,.\] Let us
interpret these two terms. The first ratio on the right-hand side provides an
estimation of the proportion $\widetilde N_{ij}/\sum_{l}\widetilde
N_{lj}$ of individuals in a site $j$ that belong to a species
$i$. This proportion is estimated by the ratio of the number $X_{ij1}$
of individuals of the species $i$ observed at site $j$ in the
opportunistic dataset to the total number $\sum_{l}X_{lj1}$ of
individuals observed at site $j$ in the same data. When the
observational  intensities $E_{j1}$ in the opportunistic dataset 1 is large,
the ratio $X_{ij1}/\sum_{l}X_{lj1}$ provides a very accurate
estimation of the abundance proportion $\widetilde
N_{ij}/\sum_{l}\widetilde N_{lj}$, and we have (see
Formula~(\ref{limit1}) in the Appendix~\ref{appendix:proof})
\begin{equation}\label{eq:explic_var}
\widehat N_{ij}\stackrel{E_{j1}\to\infty}{\approx} {\tN_{ij}\over
  \sum_{l}\tN_{lj}}\times {\sum_{l}X_{lj0}\over \tE_{j0}}\,. 
\end{equation}
The second term in the right-hand side of~(\ref{eq:explic_var}) provides an estimation of the total
(normalized) relative abundance $\sum_{l}\widetilde N_{lj}$ at the site $j$.  
This total (normalized) abundance is estimated from the dataset 0 by
dividing the total number $\sum_{l}X_{lj0}$ of individuals counted at
the site $j$ in the dataset 0 by the (normalized) observational  intensity
$\widetilde E_{j0}$. Let us now explain the reduction of variance
observed in~(\ref{variance}). 
The formula~(\ref{eq:explic_var}) shows that we estimate $\widetilde
N_{ij}$ by first estimating the total (normalized) relative abundance
$\sum_{l}\widetilde N_{lj}$ with the dataset 0 and then renormalize
this estimation with the ratio $\widetilde N_{ij}/\sum_{l}\widetilde
N_{lj}$ which has been accurately estimated with the dataset 1. The
reduction of variance observed in~(\ref{variance}) then results from
the use of  the whole
counts $\sum_{l}X_{lj0}$ at site $j$ in the dataset 0 for estimating $\widetilde N_{ij}$ instead of the
sole counts $X_{ij0}$ of the species $i$ at site $j$.

\subsection{Species not monitored in the scheme characterized by a
  known sampling observational intensity}\label{sec:notmonitored}

As already mentioned, combining the two datasets also allows to
estimate $\tN_{ij}$ for some species $i$ that are not monitored in the
dataset 0, but are monitored in the opportunistic dataset 1. This
situation formally corresponds to the case where $P_{i0}=0$. For
$E_{j1}\to\infty$, the limit variance of the estimator $\widehat N_{ij}$
is (see Formula~(\ref{var3}) in the Appendix~\ref{appendix:proof})
\[
\textrm{var}(\widehat N_{ij})\stackrel{E_{j1}\to\infty}{\sim}{\tN_{ij}^2\over \sum_{l}\tP_{l0}\tN_{lj}\tE_{j0}}.
\]
Because the species $i$ is not monitored in dataset 0, the (normalized) relative abundance
$\tN_{ij}$ cannot be estimated with the sole dataset 0.  Thus, there
is an obvious improvement to be made by using our estimation scheme
that combines the two datasets. To reveal the power of our approach,
let us compare the variance $\textrm{var}(\widehat N_{ij})$ of our
relative abundance estimator with the variance of the imaginary estimator
$\hN_{ij}^{0,\textrm{imaginary}}$ based on an imaginary dataset 0 where the
species $i$ would have been monitored with some (imaginary) detection/reporting probability
 $P^{\textrm{imaginary}}_{i0}$.  The variance of the maximum
likelihood estimator $\hN_{ij}^{0,\textrm{imaginary}}$ of $\tN_{ij}$
with this imaginary dataset 0 would be
$\tN_{ij}/(\tE_{j0}\tP_{i0}^{\textrm{imaginary}})$ so that
\[
\textrm{var}(\widehat N_{ij})\stackrel{E_{j1}\to \infty}{\sim}\textrm{var}(\widehat N_{ij}^{0,\textrm{imaginary}})\times
{P_{i0}^{\textrm{imaginary}}N_{ij}\over \sum_{l}P_{l0}N_{lj}}\,.
\]
In particular, the estimation provided by $\hN_{ij}$ can significantly
outperform the imaginary estimation we would have obtained with the sole
imaginary dataset 0 (where the species $i$ would have been
monitored). Moreover, if we compare the estimator $\widehat N_{ij}$
with the imaginary estimator $\widehat N_{ij}^{\textrm{imaginary}}$ based
on both the imaginary dataset $k=0$ and the dataset $k=1$, we observe
that the ratio of their variance
\[{\textrm{var}(\widehat N_{ij})\over \textrm{var}(\widehat N_{ij}^{\textrm{imaginary}})}={P_{i0}^{\textrm{imaginary}}N_{ij}+\sum_{l}P_{l0}N_{lj}\over \sum_{l}P_{l0}N_{lj}}\]
remains close to one when $P_{i0}^{\textrm{imaginary}}N_{ij} \ll
\sum_{l}P_{l0}N_{lj}$. This means that with our estimation scheme,
there is not much difference between the estimation based on a dataset
collected with known observational intensities where a species $i$ is rare and the
estimation based on a dataset collected with known observational intensities where a
species $i$ is not monitored. In other words, there is no instability
on the estimation of the relative abundance of a species when it is not present
in the dataset collected with known observational intensities.

\section{Illustration}\label{sec:illustration}

\subsection{Datasets}

In this section, we investigate on some datasets the predictive power
of our modeling approach. We estimated the relative abundance of 34
bird species in the non-urban habitat of 63 sites in the Aquitaine
region (South West of France).  We fitted our model with an
opportunistic dataset and a dataset collected with known observational
intensity. We then assessed the predictive power of our approach with
the help of an independent dataset collected with known observational
intensity in the same area, hereafter referred as ``validation
dataset''. We therefore illustrate the ability of our approach to
provide better predictions of species relative
abundance than other approaches based on either of the two datasets
alone.

\begin{figure}
  \centerline{%
    \includegraphics[width=15cm,keepaspectratio]{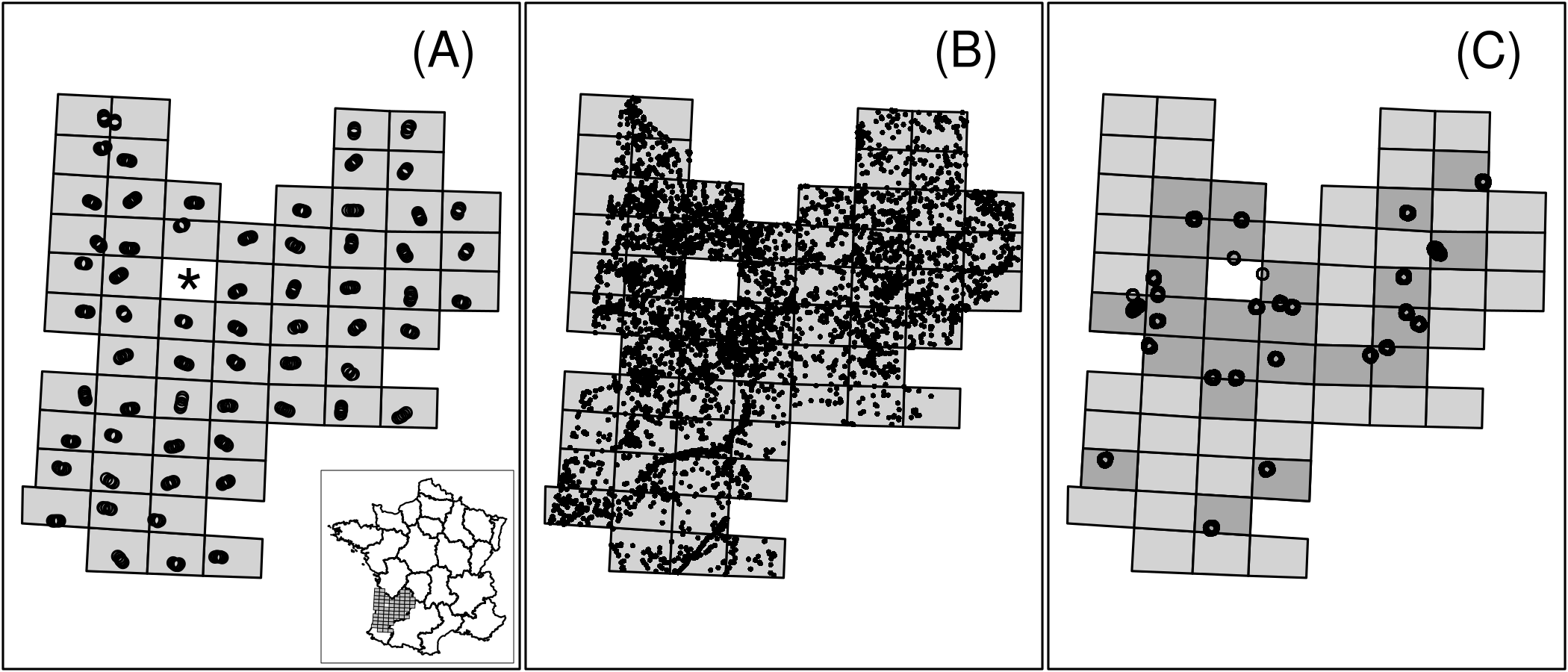}}
  \caption{The datasets used to illustrate our statistical
    framework. The location of the Aquitaine region in France is
    displayed in the insert. (A) distribution of the ACT listening
    points in the region; (B) distribution of the LPO records
    (opportunistic dataset) in the region; (C) distribution of the
    STOC listening points in the region. The grey quadrat cells are
    used as the ``sites'' in our analysis (they measure $\approx
    30\times 20$ km). Note that the quadrat cell containing the
    Bordeaux metropolitan area (indicated by an asterisk in (A)) has
    been removed from the dataset.}
  \label{fig:fig1}
\end{figure}

We first describe the opportunistic dataset. We used the recent online
database developed by the Ligue de Protection des Oiseaux (LPO, Bird
Life representative in France, largest French bird watcher NGO, with
regional delegations). This online system was launched successively by
the different regional LPO groups, and we acquired data from one of
the first groups to start, Aquitaine, South-Western France, with data
collection starting in 2007 (www.fauneaquitaine.org). Any citizen who
can identify bird species can register on this website and record any
bird observation s/he wishes, noting the species, date, and location
(to the nearest 500 m). Hundreds of observers thus record hundreds of
thousands observations. We typically ignore why these observations
were made, e.g., the motivation of the observer, the reason for
choosing to report these observations over others, whether they report
all the species they have seen at a given place and time, the
underlying observational intensity, etc. We selected all such
opportunistic records between April and mid-June 2008--2011. For each
record, we considered the number of animals detected by the
observer. Data were pooled over years, because we will focus here only
on spatial variation in relative abundance. Over 115~000 species
records detected in a non-urban habitat were considered in this study
(see Fig. \ref{fig:fig1}B).

We then describe the dataset collected with known observational
intensity, used for the fit of the model. We used the data from the
ACT monitoring plan jointly carried out by the French National Game
and Wildlife Agency (ONCFS, Office National de la Chasse et de la
Faune Sauvage), the national hunter association (FNC,
F\'{e}d\'{e}ration Nationale des Chasseurs) and the French
departemental hunters associations (FDC, F\'{e}d\'{e}rations
D\'{e}partementales des Chasseurs). The main objective of the ACT
survey was to monitor the breeding populations of several migratory
bird species in France \citep{Boutin2003}; ACT stands for
\textit{Alaudidae}, \textit{Columbidae}, \textit{Turdidae}, which were
the main bird clades of interest for this monitoring, though this
program also monitors several \textit{Corvidae} species (see table
\ref{tab:tab2} for the list of species of interest for our
study). Thus, only a fraction of the species recorded by the LPO
program was also studied by the ACT survey. The Aquitaine region was
discretized into 64 quadrat cells, and in each cell, a 4km long route
was randomly sampled in the non-urban habitat of the cell (see
Fig. \ref{fig:fig1}(A)). Each route included 5 points separated by
exactly 1 km. Each route was traveled twice between April and
mid-June, and every point was visited for exactly 10 minutes within 4
hours after sunrise in appropriate weather conditions. Every bird
heard or seen was recorded, and for each point and each species, the
maximum count among the two visits was retained. The observers were
professionals from the technical staff of either the ONCFS or the
hunters associations. Note that due to organization constraints, some
listening points in a site were not necessarily counted every
year. Between 2008 and 2011, over 9~500 birds were counted.

Finally, we describe the validation dataset, used to assess the
predictive power of our model. We used the data from the STOC program
(\textit{Suivi temporel des oiseaux communs}), a French breeding bird
survey carried out by the French museum of natural history (MNHN,
Museum National d'Histoire Naturelle) for the same region and the same
years. The STOC survey \citep{Jiguet2012} is based on a stratified
random sampling, with each volunteer observer being assigned a 2
$\times$ 2 km square randomly chosen within 10 km of his house. The
observer then homogeneously distributed 10 points within the
square. Each point was visited twice between April and mid-June
(before and after May 8th, with at least 4 weeks between visits) for
exactly 5 minutes within 4 hours after sunrise in appropriate weather
conditions (no rain or strong winds). Every bird heard or seen was
recorded, and for each point and each species, the maximum count among
the two visits was retained. These counts were then summed for a given
square, year and species. Between 2008 and 2011, 251 listening points
belonging to 29 such squares have been surveyed in non-urban habitat
(to allow the comparison with the other datasets, we removed the
listening points located in urban habitat), most of them for several
years, and over 15241 birds were detected by the observers.

Our aim was to test our model ability to provide a better prediction
of the spatial variation in species relative abundance than any model
based on either of the two datasets alone. The ``sites'' of our model
were the 63 quadrat cells defined for the ACT survey; we removed the
quadrat cell containing the metropolitan area of Bordeaux (a large
town with a population of $>$ 1 million inhabitants), where the
sampling process in the opportunistic dataset could not be supposed to
be the same as in the other areas (see Fig. \ref{fig:fig1}(A)). We
focused on $I=34$ bird species (see Table \ref{tab:tab2}). Note that
the smaller number of species monitored in the ACT survey allowed to
demonstrate the ability of our approach to estimate the relative
abundance of species monitored only in opportunistic dataset. For both
the ACT survey and the STOC survey, the observational intensity in the
site $j$ was measured as the number of points-years sampled in the
quadrat cell $j$ during the period 2008--2011. We used the validation
STOC dataset to assess the predictive power of our modeling
approach. Only 24 sites contained at least one STOC listening point
(Fig. \ref{fig:fig1}(C)), so that this assessment was restricted to
these sites.

\begin{table}
  \caption{List of the 34 bird species under study. The 13 species
    monitored only by the ACT survey are indicated by an
    asterisk. All species were surveyed by the STOC and the LPO
    program. 
  } 
  \label{tab:tab2}
  \begin{tabular}{ll}
    \hline
    Latin name & species  \\ 
    \hline
    \textit{Aegithalos caudatus} & Long-Tailed Tit \\
    \textit{Alauda arvensis}$^*$ & Eurasian Skylark  \\
    \textit{Alectoris rufa}$^*$ & Red-Legged Partridge  \\
    \textit{Carduelis carduelis} & European Goldfinch  \\
    \textit{Carduelis chloris} & European Greenfinch \\
    \textit{Certhia brachydactyla} & Short-Toed Treecreeper \\
    \textit{Columba palumbus}$^*$ &  Common Wood Pigeon  \\
    \textit{Coturnix coturnix}$^*$ & Common Quail \\
    \textit{Cuculus canorus} & Common Cuckoo  \\
    \textit{Dendrocopos major} & Great Spotted Woodpecker \\
    \textit{Erithacus rubecula} & European Robin  \\
    \textit{Fringilla coelebs} & Common Chaffinch \\
    \textit{Garrulus glandarius}$^*$ & Eurasian Jay \\
    \textit{Hippolais polyglotta} & Melodious Warbler \\
    \textit{Lullula arborea}$^*$ & Woodlark  \\
    \textit{Luscinia megarhynchos} & Common Nightingale  \\
    \textit{Milvus migrans} & Black Kite  \\
    \textit{Cyanistes caeruleus} & Eurasian Blue Tit  \\
    \textit{Parus major} & Great Tit  \\
    \textit{Passer domesticus} & House Sparrow \\
    \textit{Phasianus colchicus}$^*$ & Common Pheasant \\
    \textit{Phoenicurus ochruros} & Black Redstar \\
    \textit{Phylloscopus collybita} &  Common Chiffchaff \\
    \textit{Pica pica}$^*$ & Eurasian Magpie  \\
    \textit{Pica viridis} & Eurasian Green Woodpecker  \\
    \textit{Sitta europaea} & Eurasian Nuthatch  \\
    \textit{Streptopelia decaocto}$^*$ & Eurasian Collared Dove \\
    \textit{Streptopelia turtur}$^*$ & European Turtle Dove  \\
    \textit{Sylvia atricapilla} & Eurasian Blackcap \\
    \textit{Troglodytes troglodytes} & Eurasian Wren \\
    \textit{Turdus merula}$^*$ & common Blackbird \\
    \textit{Turdus philomelos}$^*$ & Song Thrush  \\
    \textit{Turdus viscivorus}$^*$ & Mistle Thrush \\
    \textit{Upupa epops} & Eurasian Hoopoe  \\
   \hline
  \end{tabular}
  \bigskip
\end{table}

\subsection{Comparison of the predictive power}

Let $X_{ijk}$ be the number of animals of the species $i$ detected in
the site $j$ in the dataset $k$. Let $k=a$ denote the dataset with
known observational intensity collected by the ACT survey; let
$k=\ell$ denote the opportunistic dataset collected by the LPO;
finally, let $k=s$ denote the validation dataset collected by the STOC
survey.  We compared different statistical approaches to estimate the
relative abundances of the species in the sites.

Let $\hN_{ij}^m$ be the relative abundance estimated for
the species $i$ in the site $j$ with the statistical approach
$m$. We estimated the relative abundance of each species $i$ in
each site $j$ with the following approaches:
\begin{eqnarray}
\hN_{ij}^{a} & =&  X_{ija} / \pi^a_j \label{eq:eqa}\\
\hN_{ij}^{s} & = & X_{ijs} / \pi^s_j \label{eq:eqb}\\
\hN_{ij}^{\ell 1} & = & X_{ij\ell} / S_j \label{eq:eqc}\\
\hN_{ij}^{\ell 2} & = & X_{ij\ell} / \sum_i
X_{ij}^{\ell} \label{eq:eqd}
\end{eqnarray}
where $\pi^k_j$ denotes the number of listening points of the site $j$
sampled in the dataset $k$, and $S_j$ denotes the area of the site $j$
(determined by intersecting each ACT quadrat with the Aquitaine
region). For the LPO dataset $k=\ell$, we had to account for the
site-specific unknown intensity. We estimated this intensity with two
proxies that are commonly used in such cases. First, we assumed that
observational intensity was spatially uniform so that it varied only
with quadrat cell area $S_j$ (the resulting approach is labeled $\ell
1$). Another proxy considered that the observational intensity within
a site was proportional to the total number of records across the
sites (pooled over all species; the resulting approach is labeled
$\ell 2$).

Finally, we fitted the model described in the previous sections, using
the ACT dataset $a$ as the dataset collected with known observational
intensity ($k=0$), and the LPO dataset $\ell$ as the opportunistic
dataset ($k=1$). Note that we supposed a quasi-Poisson distribution,
to account for moderate overdispersion in our dataset. Thus, we could
estimate the value of $\hN_{ij}^{\ell+a}$ with our approach.

The relative abundance is the absolute abundance multiplied by an
unknown constant, and this constant may vary among approaches.
Therefore, to allow the comparison between the various approaches, we
standardized the relative abundance estimates in the following way:
\[
\tN_{ij}^m = {\hN_{ij}^m \over \sum_{j} \hN_{ij}^m}
\]
We want to investigate whether the estimates obtained by our model
are closer or not to the true densities than any of the estimates
that could be obtained from the individual datasets. We used the value
$\tN_{ij}^s$ estimated with the validation STOC dataset as the value
of reference. We assessed the predictive power of each approach $m$ by
calculating, for each species, the Pearson correlation coefficient
between the standardized relative abundance $\tN_{ij}^m$ estimated
with the method $m$ and the standardized relative abundance
$\tN_{ij}^s$ estimated with the validation dataset. We summarized this
power by calculating the median and interquartile range (IQR) of these
coefficients over the different species of interest. Although the
relative abundance estimates were calculated on the complete dataset,
these results were presented by separating the species monitored in
the ACT survey, and the species not monitored in this survey. This
allowed to evaluate the ability of our approach to estimate the
relative abundance of species not monitored in the standardized
dataset.

We also investigated the stability of our statistical approach when the
standardized dataset is small. We therefore assessed this stability by
replacing our big standardized ACT dataset $a$ by a much smaller
dataset $a'$. We subsampled the dataset $a$: for each site, we
randomly sampled only one listening point in every site, and we
considered the bird counts of only one randomly sampled year for every
point. Thus, we artificially divided the observational intensity by 18
in average in this dataset: the complete ACT dataset $a$ stored the
bird counts carried out in 1107 listening points-years, whereas the
reduced dataset $a'$ stored the bird counts carried out in only 63
listening-points-years (one in every site). We also estimated the
standardized relative abundance $\tN_{ij}^{a'} = X_{ij}^{a'}/\sum_j
X_{ij}^{a'}$ with this reduced dataset. Finally, we estimated the
relative abundance $\tN_{ij}^{\ell+a'}$ by combining this reduced
standardized dataset with the opportunistic dataset according to our model. We also assessed
the predictive power of these two approaches by comparing the
estimates with the reference values obtained with the STOC dataset.

The online supplementary material contains the data and the code for
the R software \citep{RCoreTeam2013} that will allow the reader to
reproduce our calculations.

\subsection{Results}

We fitted our model on the LPO and ACT datasets. There was only a
small amount of overdispersion in our data (the coefficient of
overdispersion was equal to 1.22); the examination of the residuals
did not reveal any problematic pattern and the quality of the fit was
satisfying. We observe in table \ref{tab:tab1} that the predictive
power was larger for our statistical approach than for all other
approaches, whether based on the dataset $a$ or $\ell$ alone.

The predictive power of our statistical approach did not decrease much
when model was fit on the smaller standardized dataset $a'$, despite
the fact that the observational intensity in this dataset was divided
by about 20. In particular, the predictive power of our approach with
a reduced dataset remained larger than the predictive power of the
other approaches. We observe a strong positive correlation between the
estimates $\tN_{ij}^{\ell+a}$ obtained with the full standardized
dataset and the estimates $\tN_{ij}^{\ell+a'}$ obtained with the
reduced standardized dataset (median Pearson's $R$ = 0.84, IQR = 0.81
-- 0.90). This illustrates clearly the gain of precision obtained by
combining the small standardized dataset with a large amount of
opportunistic data, which we demonstrated in section
\ref{sec:vared}. The very fine-grained distribution of observations
contained in the opportunistic dataset can more efficiently predict
site-specific variation in relative abundance than can the
standardized dataset.

\begin{table}
  \caption{Predictive capabilities of the various possible approaches
    to estimate the relative abundance of 34 bird species in 63 sites
    in the Aquitaine region. For each possible estimation approach
    $m$, we present the median (calculated over the species) of the
    Pearson's correlation coefficient between the relative
    abundance $\tN_{ij}^m$ estimated by the approach $m$ and the
    relative abundance $\tN_{ij}^s$ estimated by the ``reference''
    STOC approach. In parentheses, we present the interquartile range
    of this coefficient. These quantities are calculated for the set of
    species only monitored in the ACT survey and for the
    set of species not monitored in this survey. } 
  \label{tab:tab1}
  \begin{tabular}{lcc}
    \hline
    Ratio & Species only in ACT & Species not monitored in ACT\\
    \hline
    $\tN_{ij}^{a+\ell}$  & 0.55 (0.38 -- 0.68) & 0.35 (0.19 -- 0.47)\\
    $\tN_{ij}^{a'+\ell}$ & 0.54 (0.25 -- 0.61) & 0.28 (0.08 -- 0.40) \\
    $\tN_{ij}^a$       & 0.27 (0.13 -- 0.49) & ---  \\
    $\tN_{ij}^{a'}$     & 0.06 (-0.07 -- 0.23) & ---  \\
    $\tN_{ij}^{\ell 1}$  & 0.29 (0.24 -- 0.55) & 0.11 (0.06 -- 0.22) \\
    $\tN_{ij}^{\ell 2}$  &  0.44 (0.35 -- 0.51) & 0.38 (0.13 -- 0.46) \\
    \hline
  \end{tabular}           
  \bigskip
\end{table}

We investigated the ability of our method to estimate the relative
abundance of species not monitored in the ACT survey. Note that the
between-site variance of the log relative abundance estimated with our
method $\tN_{ij}^{a+\ell}$ was larger in average for the species
monitored in the ACT survey (median = 2.41, IQR = 1.1 -- 104)
than for the species not monitored in this survey (median = 1.15, IQR
= 1.04 -- 1.34), which resulted in smaller Pearson's coefficient for
the latter species (Tab. \ref{tab:tab1}). Our approach performed
better than the approaches based on the dataset $a$ or $\ell 1$
alone. The predictive power of our approach and the approach $\ell 2$
were similar. Actually, the log observational intensity estimated in a
site by our approach for the LPO dataset was strongly correlated with
the logarithm of the total number of birds detected in this site
(Pearson's $R$ = 0.85), which supports to some extent the common
practice of biologists to use the total number of birds detected in a
place as a measure of the observational intensity.

\section{Discussion}\label{sec:discussion}

\subsection{Overview}

We propose a general approach to estimate relative abundances of
multiple species on multiple ''sites'' (corresponding to different
times and/or locations) by combining one or several datasets collected
according to some standardized protocol with one or several datasets
of opportunistic nature. The estimation is performed with the
generalized linear model~(\ref{GLM}).  This modeling relies on several
assumptions, including: (i) the datasets have the same spatiotemporal
extent, (ii) the individuals of the monitored species do not cluster
into large groups, (iii) either the habitat types are known or the
observational bias towards some habitat types are the same across the
different sites. In particular, the third hypothesis is quite
restrictive and handling cases where it is not met requires
significant additional modeling.

We have demonstrated both theoretically (under the assumption that the
model is well-specified) and numerically on some datasets, that
combining opportunistic data with standardized surveys produces more
reliable estimates of the relative abundances than either dataset
alone. In particular, we observe an improvement in our example
Section~\ref{sec:illustration} even if the above hypothesis (iii) is
probably violated. We have also shown that combining opportunistic data with
standardized data allows for estimating relative abundance for species
which are not monitored in the standardized dataset.

Our approach for combining opportunistic data with survey data is
quite general: It requires to be extended in order to overcome the
current limitations (see the discussion in the next section) and to be
adapted to the specifics of each case study. Yet, we highlight two
already promising applications of our framework.  First, we emphasize
that our framework can be readily used to estimate temporal
changes. In such cases, the ''sites'' $j$ correspond to different
times $j$ and $E_{j1}$ represent the parameters describing the unknown
observational intensity at time $j$ for the opportunistic dataset. For
temporal variation, biased attention for some habitats in the
opportunistic dataset will meet the hypothesis (iii) as long as this
biased attention is constant over time. As explained in
Section~\ref{sec:assumptions}, such biases will be entirely captured
in the estimation of the $P_{ik}$.  For example, the accuracy of bird
population trends for France will be considerably improved by the
addition of opportunistic data to the current Breeding Bird Surveys.

Another very interesting feature of our framework is its ability to
estimate the relative abundance of very rare species, even if these
species are not monitored with a scheme with known sampling
effort. This has important practical implications. For example,
\citet{Guisan2006} noted ``in a sample of 550 plots surveyed in a
random-stratified way based on the elevation, slope, and aspect of the
plot during two consecutive summers in the Swiss Alps (704.2 km$^2$),
not one occurrence of the rare and endangered plant species
\textit{Eryngium alpinum} L. was recorded. This was despite the
species being easily detectable if present and independent records of
the species existing in the area within similar vegetation types.''
Our framework would be very useful in this context. In particular, if
a citizen science program collects opportunistic data on this species
along with some other more common species, then the relative abundance
of the rare species can be estimated by combining these opportunistic
data with standardized surveys monitoring the same common species.

\subsection{Limitations and extensions}

We derived from our analysis Section~\ref{sec:count:modeling} a model
based on the Poisson distribution. In practice, we may observe some
overdispersion in the data. Causes of overdispersion include
clustering of individuals, spatial auto-correlation, identification
errors, etc. It is then wise to account for overdispersion in the
modeling (see Section~\ref{sec:illustration}).

The main assumption in our modeling~(\ref{model}) is that the
observational bias $O_{ijk}$ can be decomposed into
$O_{ijk}=P_{ik}E_{jk}$. As explained in Section~\ref{sec:assumptions},
this mainly amounts to assume that the habitat types are known or the
observational bias towards some habitat types are the same across the
different sites. This assumption will not be met in many cases and we
can expect a significant improvement by taking habitat types
heterogeneity into account. This issue requires a significant
additional modeling and it will be developed elsewhere.

In our estimation framework, we did not take into account any variable
affecting the distribution of the relative abundance in the different
sites. However, it is well-known that there might be a spatial (if the
``sites'' are spatial units) or temporal (if the ``sites'' are time
units) autocorrelation in the densities. For example, it is frequent
that if the abundance of a given species is high in a given spatial
unit, it will also be high in neighboring units. Moreover, spatial
units with a similar environmental composition will often be
characterized by similar abundances. Explicitly accounting for these
patterns in the estimation process could lead to an
increased accuracy of the estimation (by reducing the effective
number of parameters). This could be done by modeling the relative
abundances $\tN_{ij}$ as a function of environmental variables, or as a
function of spatial effects \citep[e.g. using conditional
autoregression effects in a hierarchical model,
see][]{Banerjee2004}. Alternatively, it is possible to maximize a
regularized log-likelihood, i.e. to maximize for example:
\[
  \log \mathcal{L} - \sum_{i=1}^I \sum_{j=1}^J \sum_{m=1}^J \nu \pi_{jm}
  (\tN_{ij}-\tN_{im})^2
\]
where $\mathcal{L}$ is the likelihood of the model, $\pi_{jm}$ is a
measure of ``environmental and spatial proximities'' between the unit
$j$ and the unit $m$, and $\nu$ is a positive parameter that
determines the strength of the penalty. The proximities could be of
any sort (e.g. taking the value 1 if the two spatial units are
neighbours, and 0 otherwise; inverse Euclidean distances between the
units in the space defined by the environmental variables, etc.). This
kind of regularization would reduce the number of effective parameters
in the model and thereby increase the accuracy of the estimation
\citep[for example, see][]{Malbasa2011}.   

Our statistical approach relies on the assumption that the measurement
errors (identification errors, false positive) were negligible. This
is a common assumption in this type of study, although recent studies
seem to indicate that (i) even a small number of false positives can
lead to biases in estimates \citep[][]{Royle2006}, and (ii) even
highly trained professionals may be subject to such errors
\citep[e.g.][]{McClintock2010}. As a solution to this problem,
\citet{Miller2011} proposed to combine data collected using different
approaches characterized by different probabilities of identification
errors (e.g. hear counts vs. visual counts). This approach has not yet
been thoroughly tested though, especially in the context of (relative)
abundance estimation. Taking into account measurement errors in our framework, e.g. by integrating 
 the approach of \citet{Miller2011},
still requires further study.

The detectability of a given species is not necessarily constant
across sites $j$ in the standardized dataset, as documented in the
literature \citep{Link1997, MacKenzie2002a}. This unaccounted
variation of detection probability will result into an unaccounted
variation of the observational intensity. Because the knowledge of
this intensity plays a crucial role in the fit of the model, such
errors may bias the estimates if this variation of detection probability is structured according to some exogenous variables (e.g. habitat types). Many statistical frameworks
based on a particular sampling design have been suggested to estimate
detectability, such as using mixture models based on repeated counts
\citep{Royle2004}. Further work is required to adapt such
methods to our proposed framework.

\section*{Acknowledgements}

We warmly thank the members of the CiSStats group for stimulating and
fruitful discussions on opportunistic data and related statistical
issues. We also thank Laurent Couzy and Ondine Filippi-Codaccioni for
facilitating access to the LPO-Aquitaine database. Many thanks are
also due to the coordinators of the ACT survey at the French wildlife
management organization (ONCFS), the French national hunters
association (FNC) and the departmental associations (FDC), for
allowing us to use this dataset in our study. This work was partially
supported by the Fondation Math\'ematiques Jacques Hadamard through
the grant no ANR-10-CAMP-0151-02 in the "Programme des Investissements
d'Avenir", by the Labex LMH, by the Mastodon program from CNRS, by the
CiSStats program from INRA and by the Chaire de Mod\'elisation
Math\'ematiques et Biodiversit\'e from VEOLIA-Ecole
Polytechnique-MNHN.

\label{lastpage}

\appendix






\section{Link with thinned-Poisson processes}\label{sec:thinned}

In Section~\ref{sec:count:modeling}, we described a first modeling of
the count data $X_{ijk}$ leading to our model~(\ref{model}).  In this
appendix, we explain how the model~(\ref{model}) can also be motivated
by another point of view relying on the inhomogeneous point process
\citep[IPP, see][]{Cressie1993}. Indeed, IPPs have recently been
shown to be a central approach to model species distribution in
ecology.  \citet{Aarts2012} have shown the close connections
existing between IPPs and resource selection functions, a commonly
used approach to model habitat selection by the wildlife
\citep{Boyce1999}. Moreover, IPPs have also been shown to generalize
other statistical approaches commonly used to model species
distribution, such as the MaxEnt approach \citep{Renner2013} or the
classical logistic regression \citep{Fithian2013}. We compare the IPP
with our approach in this section.

The framework of IPPs suppose that the individuals of the species $i$
are distributed on a domain $\mathcal{D}$ according to a Poisson point
process with intensity $\lambda_{i}(s)$. If we assume that the
individual at location $s$ is detected and recorded in the dataset $k$
with probability $b_{ik}(s)$, then the individuals of the species $i$
recorded in the dataset $k$ are distributed according to a Poisson
point process with intensity $\lambda_{i}(s)b_{ik}(s)$. The
multiplication of $\lambda_i(s)$ with $b_{ik}(s)$ results in a
``thinning'' of the IPP; for this reason, the resulting point process
is sometimes called thinned-Poisson process
\citep[e.g.][]{Fithian2014}. Note that in the context of IPPs,
each individual is supposed to be counted at most once in each dataset
(undercounting).  On the contrary, in Section ~\ref{sec:count:modeling}, we allowed multiple counts of a single
individual during the multiple visits in a site, which makes our
development more sensible for studies characterized by a strong
observational intensity (which is generally the case of citizen
science data).

However, even with this difference, our model~(\ref{model}) can be
motivated in the context of IPPs. We can adopt different points of
view for estimating relative abundances with this modeling based on
IPPs.  A first point of view is to introduce a model for the abundance
intensities $\lambda_{i}(s)$ and the probabilities $b_{ik}(s)$ and
then estimate these quantities accordingly.  Such a point of view has
been successfully developed in a simultaneous and independent work by
\citet{Fithian2014}: They model the abundances intensities by
$\lambda_{i}(s)=e^{\alpha_{i}+\beta_{i}^Tx(s)}$ with $x(s)$ some
observed environmental variables, the probabilities by
$b_{i1}(s)=e^{\gamma_{i}+\delta^Tz(s)}$ with $z(s)$ some other
observed environmental variable and $b_{i0}(s)=1$ at locations where
survey data are available and $b_{i0}(s)=0$, else. The abundance
intensities are then estimated by $\widehat
\lambda_{i}(s)=e^{\widehat\alpha_{i}+\widehat\beta_{i}^Tx(s)}$, with
$\widehat\alpha_{i}$ and $\widehat\beta_{i}$ some penalized maximum
likelihood estimators of $\alpha_{i}$ and $\beta_{i}$.

An alternative point of view, which corresponds to the point of
view developed in this paper, is not to try to infer the intensities
$\lambda_{i}(s)$ for each $s$, but instead, to work at the scale of a
whole site $S_{j}\subset \mathcal{D}$ and infer the mean abundance
$\Lambda_{ij}=\int_{S_{j}}\lambda_{i}(s)\,ds$ of the species $i$ on
$S_{j}$.  An important feature is that we do not model the abundance
intensities $\lambda_{i}(s)$ and the probabilities $b_{ik}(s)$ in
terms of some observed environmental variables, but rather simply
assume some structural properties on these functions. In particular,
the mean abundance $\Lambda_{ij}$ in the site ${j}$ is not assumed to
be completely driven by some observed environmental variables.

Let us explain how the model~(\ref{model}) can arise in such a
context.  Let us denote by $d_{ij}(s)=\lambda_{i}(s)/\Lambda_{ij}$ the
probability density distribution describing the probability for a
given individual of the species $i$ in the site ${j}$ to be located in
$s\in S_{j}$. The number $X_{ijk}$ of individuals of the species $i$
counted in the site ${j}$ in the dataset $k$ is then distributed
according to
\[X_{ijk}\sim\textrm{Poisson}\pa{\Lambda_{ij}O_{ijk}}\quad\textrm{with}\ \ O_{ijk}=\int_{S_{j}}d_{ij}(s)b_{ik}(s)\,ds\,.\]
Let us describe some scenarii, where the observational bias $O_{ijk}$
can be decomposed as $O_{ijk}=P_{ik}E_{jk}$, leading to the
model~(\ref{model}).

In the three examples below, we will assume that the detection/reporting probability $b_{ik}(s)$ can be decomposed in 
\begin{equation}\label{IPP:bias}
b_{ik}(s)=p_{ik}\phi_{k}(s)
\end{equation}
 with $\phi_{k}(s)$ not depending on $i$. This means that the detection/reporting bias $b_{ik}(s)/b_{1k}(s)=p_{ik}/p_{1k}$ towards the species $i$ in the dataset $k$ is independent of the location $s$ (in other words the functions $b_{1k}(s),\ldots,b_{Ik}(s)$ are proportional one to the others). When this property is met we have the decomposition
 \[O_{ijk}=p_{ik}\int_{S_{j}}d_{ij}(s)\phi_{k}(s)\,ds.\]
 The decomposition does not give a decomposition $O_{ijk}=P_{ik}E_{jk}$ in general. Yet, such a decomposition arises in the three scenarii described below (which are the counterparts of the three examples described in Section~\ref{sec:assumptions}).
 \medskip
 
\noindent {\bf Example 1: sites with homogeneous habitat type.}
 Assume that the species intensity ratios $\lambda_{i}(s)/\lambda_{i'}(s)$ depend on the species $i,i'$ and the site ${j}$, but not on the location $s\in S_{j}$. Such a property is likely to be met if the site ${j}$ has an homogeneous habitat type. In this case, we have 
 $\lambda_{i}(s)/\lambda_{i'}(s)=\Lambda_{ij}/\Lambda_{i'j}$ and hence $\lambda_{i}(s)=\Lambda_{ij}g(s)$ for all $i$ and $s\in S_{j}$. 
 Then, we have 
 \[O_{ijk}=P_{ik}E_{jk}\quad\textrm{with}\ \ P_{ik}=p_{ik}\ \ \textrm{and}\ \ E_{jk}=\int_{S_{j}}g(s)\phi_{k}(s)\, ds.\]

\noindent {\bf Example 2: observations with known habitat type.}
In this example, we assume that for each observation we know in which habitat type $h(s)$ it has occurred (in particular, it will be the case if we know the location $s$ of each observation). 
Exactly as in the Example 2 in Section~\ref{sec:assumptions}, we define $\tk$ as the couple $\tk=(h,k)$. 
Assume that the density distribution $d_{ij}(s)$ depends on the species $i$ only through the habitat $h(s)$ of $s$: For any $i,i'$ and $s,s'\in S_{j}$ such that $h(s)=h(s')$ we have $d_{ij}(s)/d_{i'j}(s)=d_{ij}(s')/d_{i'j}(s')$.  In this case, we have a decomposition
$d_{ij}(s)=\alpha_{ih(s)}g(s)$ for all $s\in S_{j}$. Let us denote by $S_{jh}=\ac{s\in S_{j}:\ h(s)=h}$ the portion of the site $S_{j}$ with habitat type $h$. For any $i,j$ and $\tk=(h,k)$, the counts $X_{ij(h,k)}$ of individuals of the species $i$ in the habitat $h$ in the site $j$ for the dataset $k$ is distributed according to
\[X_{ij(h,k)}\sim\textrm{Poisson}\pa{\Lambda_{ij}P_{i(h,k)}E_{j(h,k)}}\quad\textrm{with} \ \
P_{i(h,k)}=\alpha_{ih}p_{i(h,k)}\ \ \textrm{and}\ \
E_{j(h,k)}=\int_{S_{jh}}g(s)\phi_{(h,k)}(s)\,ds.\]
We then have the decomposition $O_{ij\tk}=P_{i\tk}E_{j\tk}$ with $\tk=(h,k)$. We emphasize that in this case the probability $p_{i(h,k)}$ appearing in the decomposition~(\ref{IPP:bias}) is allowed to depend on the habitat type $h$ (the bias towards some species may differ depending on the habitat type). 
\medskip

\noindent{\bf Example 3: homogeneous distribution of habitat types.}
We do not assume anymore that the habitat type $h(s)$ for each observation is known. We assume again that we have the decomposition $d_{ij}(s)=\alpha_{ih(s)}g(s)$ for all $s\in S_{j}$, hence
\[O_{ijk}=p_{ik}\sum_{h}\alpha_{ih}\int_{S_{jh}}g(s)\phi_{k}(s)\,ds.\]
If we assume in addition that
\begin{equation}\label{IPP:prop2bis}
\int_{S_{jh}}g(s)\phi_{k}(s)\,ds=Q_{hk}\int_{S_{j}}g(s)\phi_{k}(s)\,ds,
\end{equation}
then
\[O_{ijk}=P_{ik}E_{jk}\quad\textrm{with}\ \ P_{ik}=p_{ik}\sum_{h}\alpha_{ih}Q_{hk}\ \ \textrm{and}\ \ E_{jk}=\int_{S_{j}}g(s)\phi_{k}(s)\,ds.\]
Let us investigate when the decomposition~(\ref{IPP:prop2bis}) can be met. Assume first that $\phi_{k}(s)=\beta_{kh(s)}\gamma_{k}(s)$ where $\gamma_{k}(s)$ reflects  local fluctuations independent of the habitat type. The function $g(s)\gamma_{k}(s)$ then represents small scale fluctuations and we can expect to have
\[\int_{S}g(s)\gamma_{k}(s)\,ds\approx q_{k} |S|,\]
for $S$ large enough. It would be the case for example if $g(s)\gamma_{k}(s)$ was the outcome of a stationary process. We then have
\[{\int_{S_{jh}}g(s)\phi_{k}(s)\,ds\over \int_{S_{j}}g(s)\phi_{k}(s)\,ds}\approx {\beta_{hk}q_{k}|S_{jh}|\over \sum_{h}\beta_{hk}q_{k}|S_{jh}|}\,.\]
When the ratios $|S_{jh}|/|S_{j}|$ do not depend on $j$, the above ratio depends on $h$ and $k$ only, so~(\ref{IPP:prop2bis}) holds.  This case corresponds to sites $S_{j}$ all having a similar distribution of habitat types. This property will be met if the sites $S_{j}$ correspond to the same location at different times $j$.

\section{Mathematical proofs}\label{appendix:proof}

\subsection{Identifiability conditions}

With the notations $n_{ij}=\log(N_{ij})$, $e_{jk}=\log(E_{jk})$ and
$p_{ik}=\log(P_{ik})$, the model~(1) described in our paper can be
recast as a classical generalized linear model
\[
X_{ijk}\sim\textrm{Poisson}(\lambda_{ijk}),\quad\textrm{with}\
\log(\lambda_{ijk})=n_{ij}+e_{jk}+p_{ik}.\]
The kernel of the design
matrix associated with this linear regression has a dimension equal to
$I+J+1$. Therefore, we need $I+J+1$ constraints to ensure the
identifiability of the model.

\subsection{Properties of the estimators}

The negative log-likelihood of the parameters
$(\tN_{ij},\tE_{jk},\tP_{ik})$ is
\[\mathcal{L}=\sum_{i\in I}\sum_{j\in J}\sum_{k\in\{0,1\}}\pa{\tN_{ij}\tE_{jk}\tP_{ik}-X_{ijk}\log(\tN_{ij}\tE_{jk}\tP_{ik})+\log(X_{ijk}!)}\]
where the parameters $\{\tE_{j0},\ j\in J\}$ and $\{\tP_{i0},\ i\in I\}$ are known, $\tP_{10}=1$ and $\tP_{i1}=1$ for all $i\in I$.

To keep the mathematical analysis of the maximum likelihood estimators
comprehensible, we focus below on the case where the $\tP_{i0}$ are
known.  The maximum likelihood estimators of $\tN_{ij}$ and $\tE_{j1}$
are then the solutions of
\begin{equation}\label{MLE}
  \hN_{ij}={X_{ij0}+X_{ij1}\over \tP_{i0}\tE_{j0}+\hE_{j1}}\quad\textrm{and}\quad \hE_{j1}={X_{\#j1}\over \hN_{\#j}}\,,
\end{equation}
where $X_{\#jk}=\sum_{i}X_{ijk}$ and $\hN_{\#j}=\sum_{i} \hN_{ij}$.

We first treat the simplest case where the $\tP_{i0}$ are all equal.

\subsubsection{Case of constant ratios $P_{i0}/P_{i1}$}

We consider in this paragraph the case where $\tP_{i0}=\tP_{10}$ for
all $i\in I$. This corresponds to the case where for all the species
$i$, the detection/reporting probability ratios $P_{i0}/P_{i1}$ are the same and
equal to $P_{10}/P_{11}$.  We derive from (\ref{MLE})
\[\hN_{\#j}={X_{\#j0}+X_{\#j1}\over \tE_{j0}+\hE_{j1}}\]
and inserting this expression in the formula for $\hE_{j1}$ we obtain
$\hE_{j1}=\tE_{j0} X_{\#j1}/X_{\#j0}$. As a consequence, we obtain the
closed-form expression for $\hN_{ij}$
\begin{equation}\label{MLEclosedformula}
\hN_{ij}={X_{ij0}+X_{ij1}\over X_{\# j0}+X_{\#j1}}\times {X_{\# j0}\over \tE_{j0}}.
\end{equation}
According to the strong law of large numbers for Poisson processes, we have
\begin{equation}\label{limit1}
\hN_{ij}\stackrel{E_{j1}\to\infty}{\to} {\tN_{ij}\over \tN_{\#j}}\times {X_{\# j0}\over \tE_{j0}}
\end{equation}
and
\[\textrm{var}(\widehat N_{ij})\stackrel{E_{j1}\to\infty}{\to} \pa{\tN_{ij}\over \tN_{\#j}}^2\times {\tN_{\# j}\over \tE_{j0}}={\tN_{ij}\over \tE_{j0}}\times {N_{ij}P_{i0}\over \sum_{l}N_{lj}P_{l0}}.\]
If we estimate $\tN_{ij}$ with the sole ``known-effort'' data
$X_{ij0}$, the maximum likelihood estimator is given by
$\hN_{ij}^0=X_{ij0}/\tE_{j0}$ and its variance equals
$\textrm{var}(\hN_{ij}^0)=\tN_{ij}/\tE_{j0}$. We can then compare the
variance of $\hN_{ij}$ and $\hN_{ij}^0$
\begin{equation}\label{var1}
\textrm{var}(\widehat N_{ij})\stackrel{E_{j1}\to\infty}{\sim}\textrm{var}(\hN_{ij}^0) \times {N_{ij}P_{i0}\over \sum_{l}N_{lj}P_{l0}}.
\end{equation}

\subsubsection{Case of arbitrary ratios $P_{i0}/P_{i1}$}
We no longer assume that the $\tP_{i0}$ are all equal. In this case,
we have no closed-form formula for $\hN_{ij}$ but we can compute a
first-order expansion of $\hN_{ij}$ in terms of the inverse of
$X_{\#j1}$.

The first step is to check that $\hN_{\#j}$ is upper-bounded independently of the $X_{ij1}$. When $P_{i0}>0$ for all $i$ (which means that the same species are monitored in the datasets 0 and 1), we have from~(\ref{MLE})
\[\hN_{ij}\leq {X_{ij0}+X_{ij1}\over \min_{i}(\tP_{i0}\tE_{i0})+X_{\#j1}/\hN_{\#j}}\,.\]
Summing these inequalities we obtain the upper-bound
\[\hN_{\#j}\leq X_{\#j0}/\min_{i}(\tP_{i0}\tE_{j0})\]
which does not depend on $X_{ij1}$. The case where $P_{i0}=0$ for some $i$ can be treated similarly: splitting apart the indices in $I_{0}=\{i\in I: P_{i0}=0\}$ and those out of $I_{0}$, we get from~(\ref{MLE})
\[\hN_{\#j}\leq {\sum_{i\in I_{0}}(X_{ij0}+X_{ij1})\over X_{\#j1}/\hN_{\#j}}+{\sum_{i\notin I_{0}}(X_{ij0}+X_{ij1})\over \min_{i\notin I_{0}}(\tP_{i0}\tE_{i0})+X_{\#j1}/\hN_{\#j}}.\]
This inequality is equivalent to
\[\hN_{\#j} \pa{1-{\sum_{i\in I_{0}}(X_{ij0}+X_{ij1})\over X_{\#j1}}}\leq X_{\#j0}/\min_{i\notin I_{0}}(\tP_{i0}\tE_{j0}).\]
In the asymptotic $E_{j1}\to\infty$ we obtain the asymptotic upper-bound
\[\hN_{\#j}\leq {X_{\#j0}\over \min_{i\notin I_{0}}(\tP_{i0}\tE_{j0})}\times {\sum_{i\in I}\tN_{ij}\over \sum_{i\in I\setminus I_{0}}\tN_{ij}}.\]

Now that we have checked that $\hN_{\#j}$ is (asymptotically) upper-bounded independently of the $X_{ij1}$, we can write a first-order expansion of the formula~(\ref{MLE})
\begin{equation}\label{expansion1}
\hN_{ij}={(X_{ij0}+X_{ij1})\hN_{\#j}\over X_{\#j1}}-{(X_{ij0}+X_{ij1})\hN_{\#j}^2\tP_{i0}\tE_{j0}\over X_{\#j1}^2}+O\pa{X_{ij1}\over X_{\#j1}^3}.
\end{equation}
Summing these expansions over $i\in I$ and simplifying the expression gives
\[\hN_{\#j}={X_{\#j0}X_{\#j1}\over \tE_{j0}\sum_{l}\tP_{l0}(X_{lj0}+X_{lj1})}\pa{1+O\pa{1\over X_{\#j1}}}.\]
Plugging this formula in (\ref{expansion1}) gives
\begin{eqnarray*}
\hN_{ij}&=&{X_{ij0}+X_{ij1}\over \sum_{l}\tP_{l0}(X_{lj0}+X_{lj1})}\times {X_{\#j0}\over \tE_{j0}}\times \pa{1+O\pa{1\over X_{\#j1}}}\\
&\stackrel{E_{j1}\to\infty}{\to}&{\tN_{ij}\over \sum_{l}\tP_{l0}\tN_{lj}}\times {X_{\# j0}\over \tE_{j0}}\,,
\end{eqnarray*}
where the last limit follows again from the law of large numbers for Poisson processes. Computing the asymptotic variance when $E_{j1}\to\infty$, we find after simplification
\begin{equation}\label{var3}
\textrm{var}(\widehat N_{ij}) \stackrel{E_{j1}\to\infty}{\to} {\tN_{ij}^2\over \sum_{l}\tP_{l0}\tN_{lj}\tE_{j0}}\ =\
{\tN_{ij}\over \tP_{i0}\tE_{j0}}\times {P_{i0}N_{ij}\over
  \sum_{l}P_{l0}N_{lj}}.
  \end{equation}

  As in the previous case, we can compare this variance to the
  variance of the maximum likelihood estimator
  $\hN_{ij}^0=X_{ij0}/(\tP_{i0}\tE_{j0})$ obtained by estimating
  $\tN_{ij}$ with the sole values $X_{ij0}$. The variance of
  $\hN_{ij}^0$ being
  $\textrm{var}(\hN_{ij}^0)=\tN_{ij}/(\tP_{i0}\tE_{j0})$, we obtain
  the reduction of variance
\begin{equation}\label{var2}
\textrm{var}(\widehat N_{ij})\stackrel{E_{j1}\to\infty}{\sim}\textrm{var}(\widehat N_{ij}^0)\times {P_{i0}N_{ij}\over \sum_{l}P_{l0}N_{lj}}.
\end{equation}

\end{document}